\documentstyle{article}
\begin{document}      

\centerline {{\Large\bf Qualitative investigation of the solutions }}
\centerline {{\Large\bf to differential equations}}  
\centerline {{\large  
(Application of the skew-symmetric differential forms)}}  

\centerline {L.I. Petrova}
\centerline{{\it Moscow State University, Russia, e-mail: ptr@cs.msu.su}}
\bigskip

The presented method of investigating the solutions to differential 
equations is not new. Such an approach was developed by Cartan [1] in 
his analysis of the integrability of differential equations. 
Here this approach is outlined to demonstrate the role of 
skew-symmetric differential forms. 

The role of skew-symmetric differential forms in a qualitative 
investigation of the solutions to differential equations is conditioned 
by the fact that the mathematical apparatus of these forms enables one 
to determine the conditions of consistency for various elements of 
differential equations or for the system of differential equations. This 
enables one, for example, to define the consistency of the partial 
derivatives in the partial differential equations, the consistency of 
the differential equations in the system of differential equations, the 
conjugacy of the function derivatives and of the initial data 
derivatives in ordinary differential equations and so on. The functional 
properties of the solutions to differential equations are just depend on 
whether or not the conjugacy conditions are satisfied. 

\subsection*{Specific features of the solutions to differential 
equations} 

The basic idea of the qualitative investigation of the solutions to 
differential equations can be clarified by the example of the 
first-order partial differential equation. 

Let 
$$ F(x^i,\,u,\,p_i)=0,\quad p_i\,=\,\partial u/\partial x^i \eqno(1)$$ 
be the partial differential equation of the first order. Let us consider 
the functional relation 
$$ du\,=\,\theta\eqno(2)$$
where $\theta\,=\,p_i\,dx^i$ (the summation over repeated indices is 
implied). Here $\theta\,=\,p_i\,dx^i$ is the differential form of the 
first degree. 

The specific feature of functional relation (2) is that in 
the general case this relation turns out to be nonidentical. 

The left-hand side of this relation involves a differential, and 
the right-hand side includes the differential form  
$\theta\,=\,p_i\,dx^i$. For this relation to be identical, the 
differential form $\theta\,=\,p_i\,dx^i$ must be a differential as well 
(like the left-hand side of relation (2)), that is, it has to be a 
closed exterior differential form. To do this it requires the commutator 
$K_{ij}=\partial p_j/\partial x^i-\partial p_i/\partial x^j$ of the 
differential form $\theta $ has to vanish. 

In the general case, from equation (1) it does not follow (explicitly) 
that the derivatives $p_i\,=\,\partial u/\partial x^i $, which obey 
to the equation (and given boundary or initial conditions of the 
problem), make up a differential. Without any supplementary conditions 
the commutator of the differential form $\theta $ defined as 
$K_{ij}=\partial p_j/\partial x^i-\partial p_i/\partial x^j$ is 
not equal to zero. The form $\theta\,=\,p_i\,dx^i$ occurs to be 
unclosed and is not a differential like the left-hand side of relation 
(2). The functional relation (2) appears to be nonidentical: the 
left-hand side of this relation is a differential, but the right-hand 
side is not a differential. 

The nonidentity of functional relation (2) points to a fact 
that without additional conditions derivatives of the initial 
equation do not make up a differential. This means that the 
corresponding solution to the differential equation $u$ will not be 
a function of $x^i$. It will depend on the commutator of the form 
$\theta $, that is, it will be a functional. 

To obtain the solution that is a function (i.e., derivatives of this 
solution form a differential), it is necessary to add the closure 
condition for the form $\theta\,=\,p_idx^i$ and for the dual form 
(in the present case the functional $F$ plays a role of the form 
dual to $\theta $) [1]: 
$$\cases {dF(x^i,\,u,\,p_i)\,=\,0\cr
d(p_i\,dx^i)\,=\,0\cr}\eqno(3)$$
If we expand the differentials, we get a set of homogeneous equations 
with respect to $dx^i$ and $dp_i$ (in the $2n$-dimensional 
space -- initial and tangential):
$$\cases {\displaystyle \left ({{\partial F}\over {\partial x^i}}\,+\,
{{\partial F}\over {\partial u}}\,p_i\right )\,dx^i\,+\,
{{\partial F}\over {\partial p_i}}\,dp_i \,=\,0\cr
dp_i\,dx^i\,-\,dx^i\,dp_i\,=\,0\cr} \eqno(4)$$
The solvability conditions for this system (vanishing of the determinant 
composed of coefficients at $dx^i$, $dp_i$) have the form: 
$$
{{dx^i}\over {\partial F/\partial p_i}}\,=\,{{-dp_i}\over {\partial F/\partial x^i+p_i\partial F/\partial u}} \eqno (5)
$$
These conditions determine an integrating direction, namely, a 
pseudostructure, on which the form $\theta \,=\,p_i\,dx^i$ occurs to be 
closed one, i.e. it becomes a differential, and from relation (2) the 
identical relation is produced. If conditions (5), that may be called 
the integrability conditions, are satisfied, the derivatives constitute 
a differential $\delta u\,=\,p_idx^i\,=\,du$ (on the pseudostructure), 
and the solution becomes a function. 
Just such solutions, namely, functions on the pseudostructures 
formed by the integrating directions, are the so-called generalized 
solutions [2]. The derivatives of the generalized solution constitute 
the exterior form, which is closed on the pseudostructure. 

(If conditions (5) are not satisfied, that is, the derivatives do not 
form a differential, the solution that corresponds to such derivatives 
will depend on the differential form commutator constructed of 
derivatives. That means that the solution is a functional rather 
then a function.) 

Since the functions that are the generalized solutions 
are defined only on the pseudostructures, they have discontinuities in 
derivatives in the directions being transverse to the pseudostructures. 
The order of derivatives with discontinuities 
is equal to  the exterior form degree. If the form of zero 
degree is involved in the functional relation, the function itself, 
being a generalized solution, will have discontinuities. 

If we find the characteristics of equation (1), it appears that 
conditions (5) are the equations for  characteristics [3]. 
That is, the characteristics are examples of 
the pseudostructures on which  derivatives of the differential equation 
constitute the closed forms and the solutions turn out to be functions 
(generalized solutions). (The characteristic manifolds of equation (1) 
are the pseudostructures $\pi$ on which the form $\theta =p_idx^i$ 
becomes a closed form: $\theta _{\pi}=d u_{\pi}$). 

Here it is worth noting that coordinates of the equations for 
characteristics are not identical to independent coordinates of 
initial space on which equation (1) is defined. The transition from 
initial space to the characteristic manifold appears to be a 
{\it degenerate} transformation, namely, the determinant of the system 
of equations (4) becomes zero. The derivatives of  equation (1) are 
transformed from the tangent space to the cotangent one. 
The transition from the tangent space, where the commutator of the form 
$\theta$ is nonzero (the form is unclosed, the derivatives do not form a 
differential), to the characteristic manifold, namely, the cotangent 
space, where the commutator becomes equal to zero (the closed exterior 
form is formed, i.e. the derivatives make up a differential), is the  
example of the degenerate transformation. 

Skew-symmetric differential forms, which, in contrast to  exterior 
skew-symmetric differential forms, are defined on manifolds 
with unclosed differential forms, were considered in the author's 
work [4]. Such skew-symmetric differential forms, which were named 
evolutionary differential forms (since they possess the evolutionary 
properties) cannot be closed forms. They emerge while describing real 
processes by differential equations. The skew-symmetric differential 
form $\theta\,=\,p_i\,dx^i$, which enters into functional relation (2), 
is the example of evolutionary skew-symmetric differential forms. Since 
the evolutionary skew-symmetric differential form is unclosed, the 
relation with such differential form turns out to be nonidentical. The 
properties of such nonidentical relation just specify functional 
properties of the solutions to differential equations [4,5]. 

The partial differential equation of the first order has been analyzed, 
and the functional relation with the form of the first degree analogous 
to the evolutionary form has been considered. 

Similar functional properties have the solutions to all differential 
equations. And, if the order of the differential equation is $k$, the 
functional relation with the $k$-degree form corresponds to this 
equation. For ordinary differential equations the commutator is produced 
at the expense of the conjugacy of derivatives of the functions desired 
and those of the initial data (the dependence of the solution on the 
initial data is governed by the commutator).

In a similar manner one can also investigate  the solutions to a system 
of partial differential equations and the solutions to ordinary 
differential equations (for which the nonconjugacy of desired functions 
and initial conditions is examined). 

It can be shown that  the solutions to equations of mathematical 
physics, on which no additional external conditions are imposed, are 
functionals. The solutions prove to be exact only under realization of 
the additional requirements, namely, the conditions of degenerate 
transformations: vanishing determinants, Jacobians and so on, that 
define the integral surfaces. The characteristic manifolds, the 
envelopes of characteristics, singular points, potentials of simple and 
double layers, residues and others are the examples of such surfaces. 

Here the mention should be made of the generalized Cauchy problem when 
the initial conditions are given on some surface. The so called 
``unique" solution to the Cauchy problem, when the output derivatives 
can be determined (that is, when the determinant built of the 
expressions at these derivatives is nonzero), is a functional since the 
derivatives obtained in such a way prove to be nonconjugated, that is, 
their mixed derivatives form a commutator with nonzero value, and the 
solution depends on this commutator. 

The dependence of the solution on the commutator can lead to instability 
of the solution. Equations that do not provided with the integrability 
conditions (the conditions such as, for example, the characteristics, 
singular points, integrating factors and others) may have the unstable 
solutions. Unstable solutions appear in the case when the additional 
conditions are not realized and no exact solutions (their derivatives 
form a differential) are formed. Thus, the solutions to the equations 
of the elliptic type can be unstable. 

Investigation of nonidentical functional relations lies at the basis 
of the qualitative theory of differential equations. It is well known 
that the qualitative theory of differential equations is based on the 
analysis of unstable solutions and integrability conditions. From the 
functional relation it follows that the dependence of the solution on 
the commutator leads to instability, and the closure conditions of the 
forms constructed by derivatives are the integrability conditions. One 
can see that the problem of unstable solutions and integrability 
conditions appears, in fact, to be reduced to the question  of under 
what conditions the identical relation for the closed form is produced 
from the nonidentical relation that corresponds to the relevant 
differential equation (the relation such as (2)), the identical relation 
for the closed form is produced. In other words, whether or not the 
solutions are functionals? This is to the same question that the 
analysis of the correctness of setting the problems of mathematical 
physics is reduced. 

Here the following  should be emphasized. When the degenerate 
transformation from the initial nonidentical functional relation is 
performed, an integrable identical relation is obtained. As the result 
of integrating, one obtains a relation that contains exterior forms of 
less by one degree and which once again proves to be (in the general 
case without additional conditions) nonidentical. By integrating the 
functional relations obtained sequentially  (it is possible only under 
realization of the degenerate transformations) from the initial 
functional relation of degree $k$ one can obtain $(k+1)$ functional 
relations each involving exterior forms of one of degrees: 
$k, \,k-1, \,...0$. In particular, for the first-order partial 
differential equation it is also necessary to analyze the functional 
relation of zero degree. 

Thus, application of the  skew-symmetric differential forms allows one 
to reveal the functional properties of the solutions to differential 
equations. 

\subsection*{Analysis of field equations} 
 
Field theory is based on the conservation laws. The conservation laws 
are described by the closure conditions of the exterior differential 
forms [6]. It is evident that the solutions to the equations of field 
theory describing physical fields can be only generalized solutions, 
which correspond to closed exterior differential forms. 

The generalized solutions (i.e. solutions whose derivatives form a 
differential, namely, the closed form), can have a differential 
equation, which is subject to the additional conditions. 

Let us consider what equations are obtained in this case. 

Return to equation (1). 

Assume that the solution does not explicitly depend on $u$ and is 
resolved with respect to some variable, for example $t$, that is, it 
has the form of 
$${{\partial u}\over {\partial t}}\,+\,E(t,\,x^j,\,p_j)\,=\,0, \quad p_j\,=\,{{\partial u}\over {\partial x^j}}\eqno(6)
$$
Then integrability conditions (5) (the closure conditions of the 
differential form $\theta =p_idx^i$  and the corresponding dual form) 
can be written as (in this case $\partial F/\partial p_1=1$) 
$${{dx^j}\over {dt}}\,=\,{{\partial E}\over {\partial p_j}}, \quad
{{dp_j}\over {dt}}\,=\,-{{\partial E}\over {\partial x^j}}\eqno(7)$$

These are the characteristic relations for equation (6). As it is well 
known, the canonical relations have just such a form. 

As a result we conclude that the canonical relations are the 
characteristics of equation (6) and the integrability conditions for 
this equation. 

The canonical relations obtained from the closure condition of the 
differential form $\theta = p_idx^i$ and the corresponding dual form, 
are the examples of the identical relation of the theory of exterior 
differential forms. 

Equation (6) provided with the supplementary conditions, namely, the 
canonical relations (7), is called the Hamilton-Jacobi equation [3]. 
In other words, the equation whose derivatives obey the canonical 
relation is referred to as the Hamilton-Jacobi equation. The derivatives 
of this equation form the differential, i.e. the closed exterior 
differential form: 
$\delta u\,=\,(\partial u/\partial t)\,dt+p_j\,dx^j\,=\,-E\,dt+p_j\,dx^j\,=\,du$.

The equations of field theory belong to this type. 
$${{\partial s}\over {\partial t}}+H \left(t,\,q_j,\,{{\partial s}\over {\partial q_j}}
\right )\,=\,0,\quad
{{\partial s}\over {\partial q_j}}\,=\,p_j \eqno(8)$$
where $s$ is the field function for the action functional 
$S\,=\,\int\,L\,dt$. Here $L$ is the Lagrange function, $H$ is the 
Hamilton function: 
$H(t,\,q_j,\,p_j)\,=\,p_j\,\dot q_j-L$, $p_j\,=\,\partial L/\partial \dot q_j$.
The closed form $ds\,=-\,H\,dt\,+\,p_j\,dq_j$ (the Poincare invariant) 
corresponds to equation (8). 

The coordinates $q_j$, $p_j$ in equation (A.8) are conjugated ones. 
They obey the canonical relations. 

In quantum mechanics (where to the coordinates $q_j$, $p_j$ the 
operators are assigned) the Schr\H{o}dinger equation serves as an analog 
to equation (8), and the Heisenberg equation serves as an analog to the 
relevant equation for the canonical relation integral. Whereas the 
closed exterior differential form of zero degree (the analog to the 
Poincare invariant) corresponds to the Schr\H{o}dinger equation, the 
closed dual form corresponds to the Heisenberg equation. 

A peculiarity of the degenerate transformation can be considered by the  
example of the field equation. The transition from the unclosed 
differential form (which is included into the functional relation) to 
the closed form is the degenerate transformation. Under degenerate  
transformation the transition from the initial manifold (on which the 
differential equation is defined) to the characteristic (integral) 
manifold goes on. 

Here the degenerate transformation is a transition from the Lagrange 
function to the Hamilton function. The equation for the Lagrange 
function, that is the Euler variational equation, was obtained from the 
condition $\delta S\,=\,0$, where $S$ is the action functional. In the 
real case, when forces are nonpotential or couplings are nonholonomic, 
the quantity $\delta S$ is not a closed form, that is, 
$d\,\delta S\,\neq \,0$. But the Hamilton function is obtained from the 
condition $d\,\delta S\,=\,0$ which is the closure condition for the 
form $\delta S$. The transition from the Lagrange function $L$ to the 
Hamilton function $H$ (the transition from variables $q_j,\,\dot q_j$ 
to variables $q_j,\,p_j=\partial L/\partial \dot q_j$) is a transition 
from the tangent space, where the form is unclosed, to the cotangent 
space with a closed form. One can see that this transition is 
a degenerate one. 

The invariant field theories used only nondegenerate transformations 
that conserve a differential. By the example of the canonical relations 
it is possible to show that nondegenerate and degenerate transformations  
are connected. The canonical relations in the invariant field 
theory correspond to nondegenerate tangent transformations. At the same 
time, the canonical relations coincide  with the characteristic relation 
for  equation (8), which the degenerate transformations correspond to. 
The degenerate transformation is a transition from the tangent space 
($q_j,\,\dot q_j)$) to the cotangent (characteristic) manifold 
($q_j,\,p_j$). On the other hand, the nondegenerate transformation is a 
transition from one characteristic manifold ($q_j,\,p_j$) to another 
characteristic manifold ($Q_j,\,P_j$). \{The formula of canonical 
transformation can be written as $p_jdq_j=P_jdQ_j+dW$, where $W$ is the 
generating function\}. 

It may be easily shown that such a property of duality is also a 
specific feature of transformations such as tangent, gradient, contact, 
gauge, conform mapping, and others. 

\bigskip

Thus, the application of mathematical apparatus of the 
skew-symmetric differential forms to qualitative investigation 
of the solutions to differential equations enables one to understand 
the specific features of the solutions to differential equations. Such 
investigation is of interest in the analysis of the equations of 
mathematical physics.

1. Cartan E., Les Systemes Differentials Exterieus ef Leurs Application 
Geometriques. -Paris, Hermann, 1945. 

2. Vladimirov V.~S., Equations of mathematical physics. -Moscow, 
Nauka, 1988 (in Russian). 

3. Smirnov V.~I., A course of higher mathematics. -Moscow, 
Tech.~Theor.~Lit. 1957, V.~4 (in Russian). 

4. Petrova L.~I., Invariant and evolutionary properties of the 
skew-symmetric differential forms. 

http://arXiv.org/pdf/math.GM/0401039

5. Petrova L.~I., Identical and nonidentical relations. Nondegenerate 
and degenerate transformations. (Properties of the 
skew-symmetric differential forms). 

http://arXiv.org/pdf/math.GM/0404109

6. Petrova L.~I., Conservation laws. Their role in evolutionary 
processes. (The method of skew-symmetric differential forms). 

http://arXiv.org/pdf/math-ph/0311008

\end{document}